\documentclass[11pt]{article}

\usepackage{booktabs}       
\usepackage{nicefrac}       
\usepackage{microtype}      
\usepackage{lipsum}
\usepackage{blindtext}

\RequirePackage{amsthm,amsmath,amsfonts,amssymb}
\RequirePackage[authoryear]{natbib}
\RequirePackage[colorlinks,citecolor=blue,urlcolor=blue,breaklinks=true]{hyperref}
\RequirePackage{graphicx}
\usepackage{breakcites}
\usepackage{siunitx}
\usepackage{multirow}
\usepackage{comment}
\usepackage[a4paper, total={6in, 9in}]{geometry}

\title{A joint modeling approach to treatment effects estimation with unmeasured confounders}

\author{
 Namhwa Lee \\
  Department of Statistics\\
  University of California Riverside\\
  Riverside, CA 92507 \\
  \texttt{nlee098@ucr.edu} \\
   $\And$
 Shujie Ma \\
  Department of Statistics\\
  University of California Riverside\\
  Riverside, CA 92507 \\
  \texttt{shujie.ma0@ucr.edu}
}

\begin{document}

\maketitle

\begin{abstract}
Estimating treatment effects using observation data often relies on the assumption of no unmeasured confounders. However, unmeasured confounding variables may exist in many real-world problems. It can lead to a biased estimation without incorporating the unmeasured confounding effect. To address this problem, this paper proposes a new mixed-effects joint modeling approach to identifying and estimating the OR function and the PS function in the presence of unmeasured confounders in longitudinal data settings. As a result, we can obtain the estimators of the average treatment effect and heterogeneous treatment effects. In our proposed setting, we allow interaction effects of the treatment and unmeasured confounders on the outcome. Moreover, we propose a new Laplacian-variant EM algorithm to estimate the parameters in the joint models. We apply the method to a real-world application from the CitieS-Health Barcelona Panel Study, in which we study the effect of short-term air pollution exposure on mental health. 

\medskip


\end{abstract}


\section{Introduction}


A primary objective of causal inference is to estimate treatment effects using observational studies, especially when randomized clinical trials are infeasible. Average treatment effects (ATEs) and heterogeneous treatment effects (HTEs) are estimated based on a set of assumptions: Consistency, Positivity, and No unmeasured confounders. Among them, the no unmeasured confounders assumption posits that potential outcomes become conditionally independent of treatment assignments given the observed confounders \citep{rosenbaum1983central}. Conventional statistical methods for treatment effects estimation often rely on this assumption, i.e., all relevant confounders are measured and included in the model \citep{greenland1986identifiability,pearl2009causality,hernan2024causal}. However, this assumption is frequently violated in observational studies, where some key confounders may be missing or imperfectly measured in the dataset \citep{mccandless2007bayesian,uddin2016methods}. Failure to account for these unmeasured confounders can result in biased treatment effect estimates, undermining the validity of causal conclusions \citep{mickey1989impact,lipsitch2010negative,vander2011new,vanderweele2019principles}.

How to address or alleviate this strong assumption has drawn increasing attention among researchers in recent years. Specifically, \cite{zhang2018addressing} and \cite{zhang2020assessing} provide a comprehensive review and discuss multiple statistical methodologies designed for this problem. They recommend using the E-value \citep{vanderweele2017sensitivity} to implement sensitivity analyses when unmeasured confounders may impact causal treatment effect estimates. \cite{streeter2017adjusting} investigates methods used to account for unmeasured confounding in longitudinal data. They conclude that several commonly employed methods include the instrumental variable method
\citep[e.g.][]{angrist1996identification}, the difference-in-differences \citep[e.g.][]{abadie2005semiparametric,imbens2009recent} approach, and the fixed effects or mixed-effects regression models \citep[e.g.][]{yang2018propensity,imai2019should,sakaguchi2020estimation}.


In this paper, we propose a new mixed-effect joint modeling approach to identifying and estimating the outcome regression (OR) function and propensity score (PS) function in the presence of unmeasured confounders in longitudinal studies. As a result, we can obtain the estimators for ATEs and HTEs. To incorporate the effect of unmeasured confounders for causality study, the mixed-effects models become an effective solution and popularly used models for the OR and PS functions, in which the unmeasured confounders are treated as the random-effect term satisfying certain distributional assumptions; see for example,    \cite{yang2018propensity}, \cite{Chang2022}, \cite{Langworthy2023}, and so forth. 
In the existing works, they use a conventional mixed-effect model for the OR function and/or the PS function. Then they apply a standard EM algorithm to estimate the mixed-effect models for the two functions, separately. Moreover, their considered models do not allow possible interaction effects between the unmeasured confounders and the treatment variable. 

Different from the existing works, we propose a new mixed-effect model for the OR function that allows the treatment variable to interact with both observed and unobserved confounders, so the coefficients for the confounders are different for the treated and untreated groups.
This poses a great technical challenge in estimating the model parameters, and the standard EM algorithm can no longer be applied in our considered scenario. Due to the interaction effect of the treatment variable and the unmeasured confounders, the conditional distribution of the unmeasured confounders in the EM algorithm is involved with the conditional distribution of the treatment variable given the observed and unobserved confounders, which is the PS function. We model the PS function by a logistic mixed-effect model and estimate the parameters in the two models for the OR and PS functions jointly.   Moreover, we allow the coefficients of the unmeasured confounders in the mixed-effects models for the OR and PS functions to be different and unknown. We propose a new Laplacian-variant EM algorithm to estimate the parameters and discuss its convergence property. It is worth noting that for sensitive analysis, several works \citep[e.g.][]{Carnegie2016,zhangsmall2020} also consider the unmeasured confounder as a random-effect term in the models for OR and PS functions. However, they pre-specify the values for the coefficients of the unmeasured confounder as well as the parameters in its distribution before fitting the model using the observed data and treat the prespecified parameters as sensitivity parameters. Their goal is to check how sensitive the conclusion about the causality is to different chosen values of the sensitivity parameters. In our considered framework, we treat all parameters in the model as unknown and estimate the unknown parameters using our proposed Laplacian-variant EM algorithm.

We apply the proposed method to the CitieS-Health Barcelona Panel Study \citep{gignac_florence_2022_6503022} to investigate the causal effects of air pollution on cognitive health. The study provides a rich dataset on environmental exposures and cognitive outcomes. 
Given the substantial evidence linking air pollution, particularly fine particulate matter (PM2.5), to adverse health outcomes, it is crucial to investigate its effects on cognitive health. Air pollution has been consistently associated with a range of negative health outcomes, including respiratory and cardiovascular diseases, as well as cognitive decline \citep{pope2002lung,weuve2012exposure,CALDERONGARCIDUENAS2015157}. Studying the impacts of air pollution on cognition has received increasing attention among researchers in recent years, especially as the global population is aging and the burden of cognitive disorders grows \citep{peters2015air,Livingston2017Dementia}. More recently, \cite{khan2019environmental} identified a significant association between air pollution and neuropsychiatric disorders, and \cite{ju2023causal} examined the causal effect of air pollution on mental health in China. 
In this paper, we apply the newly proposed joint mixed-effect modeling method to investigate the causal effect of air pollution exposure on cognitive health. Our study provides new insights into the cognitive impacts of pollution, as we obtain the estimates for not only ATEs but also HTEs, so that we can see how the causal effects of air pollution on cognition can change with the values of the confounders, which have important implications for environmental policy and public health interventions aimed at reducing exposure to harmful pollutants. Moreover, we conduct the casual inference by incorporating possible unmeasured confounders and perform a sensitivity analysis based on our model to show that the unmeasured confounders play a significant role in the causality study of pollution, so they cannot be ignored in the procedure of treatment effect estimation. 

The rest of the paper is organized as follows. Section \ref{setup} introduces the basic model setup and the identification of causal effects. Section \ref{estimation} introduces our method and the proposed Laplacian Variant EM algorithm. In Section \ref{numerical}, we conduct numerical studies. We illustrate the proposed method by simulation studies and also apply the method to the CitieS-Health Barcelona Panel Study \citep{gignac_florence_2022_6503022} to investigate the causal effect of air pollution on cognitive outcomes. 

\section{Set  up and model formulation}\label{setup}

\subsection{Set up}

In this study, we assume that each subject ($i = 1, 2, \ldots, m$) has $n_{i}$ repeated measurements. For each individual $i$, we have a set of variables $\{ Y_{ij}, D_{ij}, \boldsymbol{Z}_{i}, \boldsymbol{X}_{ij}, U_{i} \}$ across $j = 1, 2, \ldots, n_{i}$ time periods, where $Y_{ij}$ denotes the observed outcome variable, and $D_{ij}$ is a binary treatment assignment indicator, taking the value 1 if treated and 0 if untreated. We divide the observed confounding variables into two parts: time-invariant ($\boldsymbol{Z}_{i} \in \mathbb{R}^{p_{1}}$) and time-variant ($\boldsymbol{X}_{ij} \in \mathbb{R}^{p_{2}})$. Additionally, we assume that each subject possesses an integrated unmeasured confounding $U_{i} \in \mathbb{R}$ which can be regarded as the linear combination of all possible unmeasured confounding factors.

We adopt the potential outcome framework to infer causal effects in observational studies, following the approach introduced by \cite{rubin1974estimating}. For subject $i$ at time point $j$, let $Y_{ij}(1)$ and $Y_{ij}(0)$ represent the potential outcomes with and without exposure to the treatment, respectively. The treatment effect for the $i$-th individual at time $j$ is then defined as the difference between these two potential outcomes, $Y_{ij}(1) - Y_{ij}(0)$. The average treatment effect (ATE) is given by the overall expectation of these differences, $\tau := \mathbb{E} \big[ Y_{ij}(1) - Y_{ij}(0) \big]$. In the context of the air pollution and mental health study, the ATE allows us to examine the overall short-term impact of air pollution on mental health test performance. Furthermore, estimating the heterogeneous treatment effect (HTE) enables us to explore the impact of air pollution on mental health across different subgroups, such as females or individuals of specific ages or exposure times to noise.

Throughout this paper, we make the following assumptions:

\begin{itemize}
   \item [$(A1)$] Conditional Ignorabillity: 
   $Y_{ij} (0), Y_{ij} (1) \perp D_{ij} \big| \boldsymbol{X}_{ij}, \boldsymbol{Z}_{i}, U_{i}$.
   \label{A1}
   \item [$(A2)$] Positivity: $0 < P( D_{ij} = 1 | \boldsymbol{X}_{ij}, \boldsymbol{Z}_{i}, U_{i}) < 1$.
   \label{A2}
   \item [$(A3)$] Consistency: $Y_{ij} = D_{ij} Y_{ij} (1) + (1 - D_{ij}) Y_{ij} (0)$.
   \label{A3}
\end{itemize}

Unlike the conventional ones which do not account for the unmeasured effect $U_{i}$, both $(A1)$ and $(A2)$ incorporate $U_{i}$, resulting in a more relaxed set of assumptions.

\subsection{Model formulation}

In this subsection, we describe the required model formulations to estimate the identified treatment effects. Under assumptions ($A1$) - ($A3$), the ATE$(\tau)$ can be identified as follows:
\begin{align*}
    \tau 
    & = \mathbb{E} \big[ Y_{ij}(1) - Y_{ij}(0) \big] \\
    & = \mathbb{E} \Big[ 
    \mathbb{E} \big[ Y_{ij} (1) - Y_{ij} (0) \big| \boldsymbol{X}_{ij}, \boldsymbol{Z}_{i}, U_{i} \big]
    \Big] \\
    & = \mathbb{E} \Big[ 
    \mathbb{E} \big[ Y_{ij} (1) \big| \boldsymbol{X}_{ij}, \boldsymbol{Z}_{i}, U_{i} \big]
    \Big] 
    - \mathbb{E} \Big[ 
    \mathbb{E} \big[ Y_{ij} (0) \big| \boldsymbol{X}_{ij}, \boldsymbol{Z}_{i}, U_{i} \big]
    \Big] \\
    & = \mathbb{E} \Big[ 
    \mathbb{E} \big[ Y_{ij} (1) \big| \boldsymbol{X}_{ij}, \boldsymbol{Z}_{i}, U_{i}, D_{ij} = 1 \big]
    \Big] 
    - \mathbb{E} \Big[ 
    \mathbb{E} \big[ Y_{ij} (0) \big| \boldsymbol{X}_{ij}, \boldsymbol{Z}_{i}, U_{i}, D_{ij} = 0 \big]
    \Big] \\
    & = \mathbb{E} \Big[ 
    \mathbb{E} \big[ Y_{ij} \big| \boldsymbol{X}_{ij}, \boldsymbol{Z}_{i}, U_{i}, D_{ij} = 1 \big]
    \Big] 
    - \mathbb{E} \Big[ 
    \mathbb{E} \big[ Y_{ij} \big| \boldsymbol{X}_{ij}, \boldsymbol{Z}_{i}, U_{i}, D_{ij} = 0 \big]
    \Big].
\end{align*}
The second equality holds due to the law of total expectation. The fourth equality holds due to the assumption $(A1)$, and the last equality holds due to the assumption $(A3)$. Hence, postulating an outcome regression model is necessary to estimate the ATE. In this study, we construct the following outcome regression model, which allows for interactions between treatment assignments and both observed and unobserved confounding factors. For $i = 1, \ldots, m$, and $j = 1, \ldots, n_{i}$, we have
\begin{align}
    E (Y_{ij} | D_{ij}, \boldsymbol{X}_{ij}, \boldsymbol{Z}_{i}, U_{i})
    & = \begin{pmatrix} 1 & D_{ij} \end{pmatrix}
    \begin{pmatrix} \boldsymbol{\beta}_{1}^{\top} \\ \boldsymbol{\beta}_{2}^{\top} \end{pmatrix} \boldsymbol{X}_{ij}^{*}
    + \begin{pmatrix} 1 & D_{ij} \end{pmatrix}
    \begin{pmatrix} \alpha_{1} \\ \alpha_{2} \end{pmatrix} 
    U_{i} 
    \label{Eq:1} \\
    & = \boldsymbol{\beta}_{1}^{\top} \boldsymbol{X}_{ij}^{*}
    + D_{ij} \boldsymbol{\beta}_{2}^{\top} \boldsymbol{X}_{ij}^{*}
    + \alpha_{1} U_{i} + D_{ij} \alpha_{2} U_{i}  \nonumber
\end{align}
where $\boldsymbol{X}_{ij}^{*} = ( 1, \boldsymbol{Z}_{i}^{\top}, \boldsymbol{X}_{ij}^{\top})^{\top} \in \mathbb{R}^{p_{1}+p_{2}+1}$ is a vector of observed covariates including an intercept term. There are two vectors of regression coefficients: $\boldsymbol{\beta}_{1} \in \mathbb{R}^{p_{1}+p_{2}+1}$, which contains an intercept and the main effects of observed covariates, and $\boldsymbol{\beta}_{2} \in \mathbb{R}^{p_{1}+p_{2}+1}$, which includes coefficients for the treatment variable and interaction effects between treatment and observed covariates. By introducing the interaction between $D_{ij}$ and $U_{i}$, this model accounts for situations where the effect of unmeasured confounders on the outcome varies depending on the treatment assignment. This approach allows our model to accommodate scenarios in which both observed and unobserved confounders affect outcome variables differently depending on the treatment assignment.

For identification purposes, we reparameterize equation (\ref{Eq:1}) by defining $b_{i}:=\alpha_{1} U_{i}$, and $\omega = \alpha_{2}/\alpha_{1}$. The model then becomes:
\begin{align}
    E ( Y_{ij} | D_{ij}, b_{i}, \boldsymbol{Z}_{i}, \boldsymbol{X}_{ij})
    & = \boldsymbol{\beta}_{1}^{\top} \boldsymbol{X}_{ij}^{*}
    + D_{ij} \boldsymbol{\beta}_{2}^{\top} \boldsymbol{X}_{ij}^{*} + \big( 1 + \omega D_{ij} \big) b_{i}.
    \label{Eq:2}
\end{align}
In the reparameterized model (\ref{Eq:2}), $b_{i}$ represents the combined effect of unmeasured confounding factors and $\omega$ indicates that the impact of $b_{i}$ on the outcome may vary depending on the treatment assignment. Based on (\ref{Eq:2}), we propose the following three-stage models for each individual unit $i$ at the $j$-th repeated measurement. First, the OR model is specified as:
\begin{align}
    Y_{ij}
    & = 
    \boldsymbol{\beta}_{1}^{\top} \boldsymbol{X}_{ij}^{*}
    + D_{ij} \boldsymbol{\beta}_{2}^{\top} \boldsymbol{X}_{ij}^{*} + \big( 1 + \omega D_{ij} \big) b_{i} + \epsilon_{ij}
    \label{Eq:3}
\end{align}
where $\epsilon_{ij}$ is normally distributed with zero mean and variance $\sigma^{2}$. Next, we assume a logistic regression model for the treatment assignment:
\begin{align}
    P( D_{ij} = 1 | \boldsymbol{X}_{ij}, \boldsymbol{Z}_{i}, b_{i})
    = \frac{\exp(\boldsymbol{\eta}^{\top} \boldsymbol{X}_{ij}^{*} + \xi b_{i})} {1+\exp (\boldsymbol{\eta}^{\top} \boldsymbol{X}_{ij}^{*} + \xi b_{i})}
    \label{Eq:4}
\end{align}
where $\boldsymbol{\eta} \in \mathbb{R}^{p_{1}+p_{2}+1}$ and $\xi \in \mathbb{R}$ represent regression coefficients for observed covariates and the integrated unobserved effect, respectively. The coefficient $\xi$ accounts for the varying effects of unmeasured confounding on both the outcome and the treatment assignment. In this stage, we further assume that $D_{ij}$ and $D_{ij'}$ are conditionally independent given $(\boldsymbol{X}_{ij}, \boldsymbol{Z}_{i}, b_{i})$, for $j \not= j'$. In the third stage, the integrated effect of unmeasured confounding ($b_{i}$) is assumed to be normally distributed with zero mean and variance $\sigma_{b}^{2}$. 
\begin{align}
    b_{i} \overset{iid} \sim N(0, \sigma_{b}^{2})
    \label{Eq:5}
\end{align}
Additionally, this effect is assumed to be independent with $\epsilon_{ij}$, as well as the observed covariates $(\boldsymbol{X}_{ij}, \boldsymbol{Z}_{i})$. We call the above models for OR and PS functions as joint mixed-effect models. 

\subsection{Treatment effects}

Under the joint mixed-effect models (\ref{Eq:3})-(\ref{Eq:5}) and assumptions ($A1$) - ($A3$), the conditional expectation of potential outcomes $Y_{ij} (d)$ given $\{ D_{ij}=d, \boldsymbol{X}_{ij}, \boldsymbol{Z}_{i}, b_{i} \}$ for $d=0,1$ can be estimated as follow:
\begin{align*}
    \mathbb{E} \left[ 
    Y_{ij} (1) \Big| D_{ij}=1, \boldsymbol{X}_{ij}, \boldsymbol{Z}_{i}, b_{i}
    \right]
    & = \big( \boldsymbol{\beta}_{1}^{\top} + \boldsymbol{\beta}_{2}^{\top} \big) \boldsymbol{X}_{ij}^{*}
    + (1 + \omega ) b_{i}, \\
    \mathbb{E} \left[ 
    Y_{ij} (0) \Big| D_{ij}=0, \boldsymbol{X}_{ij}, \boldsymbol{Z}_{i}, b_{i}
    \right]
    & = \boldsymbol{\beta}_{1}^{\top} \boldsymbol{X}_{ij}^{*} + b_{i}.
\end{align*}
This leads to the heterogeneous treatment effect (HTE) given both observed and integrated unobserved confounding factors:
\begin{align}
    \mathbb{E} \left[ Y_{ij} (1) - Y_{ij} (0) \big| b_{i}, \boldsymbol{X}_{ij}, \boldsymbol{Z}_{i} \right]
    = \boldsymbol{\beta}_{2}^{\top} \boldsymbol{X}_{ij}^{*}
    + \omega b_{i}.
    \label{Eq:6}
\end{align}
Next, integrating out (\ref{Eq:6}) with respect to $b_{i}$ results in the HTE given the observed covariates:
\begin{align}
    \mathbb{E} \left[ Y_{ij} (1) - Y_{ij} (0) \big| \boldsymbol{X}_{ij}, \boldsymbol{Z}_{i} \right]
    = \boldsymbol{\beta}_{2}^{\top} \boldsymbol{X}_{ij}^{*}
    \label{Eq:7}
\end{align}
Finally, the ATE can be obtained by integrating (\ref{Eq:7}) with respect to the observed covariates:
\begin{align}
    \mathbb{E} \big[
    Y_{ij} (1) - Y_{ij} (0)
    \big]
    =  \boldsymbol{\beta}_{2}^{\top} \mathbb{E} \big[ \boldsymbol{X}_{ij}^{*} \big]
    \label{Eq:8}
\end{align}
The estimation procedure for model parameters will be described in Section 3.

\section{Estimation}\label{estimation}

\subsection{The EM Algorithm}

As detailed in Section 2, we formulate a three-stage model assuming that we have observations $\{ y_{ij}, d_{ij}, \boldsymbol{z}_{i}, \boldsymbol{x}_{ij}, b_{i} \}$ for each individual across the repeated measurements. However, due to the unobservable nature of $b_{i}$, the actual observations comprise only $\{ y_{ij}, d_{ij}, \boldsymbol{z}_{i}, \boldsymbol{x}_{ij} \}$. This naturally leads to an incomplete data problem involving latent or missing variables, $b_{i}$. The challenge, therefore, lies in estimating the model parameters $\boldsymbol{\theta} := (\boldsymbol{\beta}, \sigma, \omega, \boldsymbol{\eta}, \xi, \sigma_{b})$ using this incomplete data. To address this challenge, we employ the Expectation-Maximization (EM) algorithm, a widely used method for computing the maximum likelihood estimates from incomplete data problems \citep{dempster1977maximum}. Various EM-type algorithms have been developed for this purpose. For instance, \cite{laird1982random} and \cite{steele1996modified} established the use of EM algorithms in estimating generalized linear mixed effect models. Similarly, we opt to employ the EM algorithm to obtain maximum likelihood estimates of each parameter in $\boldsymbol{\theta}$ while considering $b_{i}$ as latent. However, in our study, we propose a novel EM algorithm because the classical EM algorithm cannot be directly applied to our specific model. Details of this novel algorithm will be described in Section 3.2.


\subsection{Laplacian-Variant EM Algorithm}

For the sake of simplification, we utilize vector notations throughout the estimation procedure. Thus, for each individual $i$, the complete dataset is represented as $\{ \boldsymbol{y}_{i}, \boldsymbol{d}_{i}, \boldsymbol{x}_{i}, \boldsymbol{z}_{i}, b_{i}\}$, while the incomplete dataset becomes $\{\boldsymbol{y}_{i}, \boldsymbol{d}_{i}, \boldsymbol{x}_{i}, \boldsymbol{z}_{i}\}$ with $b_{i}$ treated as latent. Lastly, the OR (\ref{Eq:3}) can be rephrased as follows:
\begin{align*}
    \boldsymbol{Y}_{i}
    & = 
    \tilde{\boldsymbol{X}}_{i} \boldsymbol{\beta}
    + 
    \big(
    \boldsymbol{1} + \omega \boldsymbol{D}_{i}
    \big)
    b_{i} + \boldsymbol{\epsilon}_{i},
\end{align*}
where $\boldsymbol{1}$ represents a $n_{i} \times 1$ vector consisting of 1, $ \boldsymbol{\beta} = (\boldsymbol{\beta}_{1}^{\top}, \boldsymbol{\beta}_{2}^{\top})^{\top} \in \mathbb{R}^{2(p_{1}+p_{2}+1)}$, $\tilde{\boldsymbol{X}}_{i}$ is a $n_{i} \times 2(p_{2}+p_{2}+1)$ design matrix constituted of $\tilde{\boldsymbol{X}}_{i} = \big( \boldsymbol{X}_{i}^{*}, \ \boldsymbol{D}_{i} [\circ] \boldsymbol{X}_{i}^{*} \big)$, and $\boldsymbol{\epsilon}_{i}$ is distributed as normal with mean $\boldsymbol{0}$ and covariance matrix $\sigma^{2} \boldsymbol{I}$ where $\boldsymbol{I}$ is an identity matrix. The EM algorithm consists of two main steps: Expectation (E) and Maximization (M).


\subsubsection{E-step} 

In the E-step, a conditional expectation of the complete data likelihood given the observed data and the current estimate of the parameters is determined \citep{wu1983convergence}. In our setup, the complete data log-likelihood is expressed as:
\begin{align}
    \ell_{c} (\boldsymbol{\theta})
    & : = \sum_{i=1}^{m} \log f(\boldsymbol{y}_{i}, \boldsymbol{d}_{i}, b_{i}, \boldsymbol{x}_{i}, \boldsymbol{z}_{i} ; \boldsymbol{\theta})
    \label{Eq:9}
\end{align}
where $\boldsymbol{\theta}$ represents a set of parameters, $(\boldsymbol{\beta}, \sigma, \omega, \boldsymbol{\eta}, \xi, \sigma_{b})$. Let $\boldsymbol{\theta}^{(k)}$ be the current estimate of these parameters. The objective function constructed in the E-step is then defined as:
\begin{align}
    Q \big( \boldsymbol{\theta} ; \boldsymbol{\theta}^{(k)} \big)
    := \mathbb{E} 
    \big[ \ell_{c} (\boldsymbol{\theta}) | \boldsymbol{y}, \boldsymbol{d}, \boldsymbol{x}, \boldsymbol{z}, \boldsymbol{\theta}^{(k)} \big].
    \label{Eq:10}
\end{align}

Given that the postulated model consists of three stages, we can decompose \eqref{Eq:10} into following three components where $\boldsymbol{\theta}_{1} = (\boldsymbol{\beta}, \sigma^{2}, \omega)$, $\boldsymbol{\theta}_{2} = (\boldsymbol{\eta}, \xi)$, and $\theta_{3} = \sigma_{b}^{2}$. 
\begin{align*}
    Q_{1} \big( \boldsymbol{\theta}_{1} ; \boldsymbol{\theta}^{(k)} \big)
    & = \mathbb{E}
    \left[
    \sum_{i=1}^{m} 
    \log f(\boldsymbol{y}_{i} | \boldsymbol{d}_{i}, b_{i}, \boldsymbol{x}_{i}, \boldsymbol{z}_{i})
    \bigg| \ 
    \boldsymbol{y}, \boldsymbol{d}, \boldsymbol{x}, \boldsymbol{z}, \boldsymbol{\theta}^{(k)}
    \right], \\
    Q_{2} \big( \boldsymbol{\theta}_{2} ; \boldsymbol{\theta}^{(k)} \big)
    & = \mathbb{E}
    \left[
    \sum_{i=1}^{m} 
    \log f(\boldsymbol{d}_{i} | \boldsymbol{x}_{i}, \boldsymbol{z}_{i}, b_{i})
    \bigg| \ 
    \boldsymbol{y}, \boldsymbol{d}, \boldsymbol{x}, \boldsymbol{z}, \boldsymbol{\theta}^{(k)}
    \right], \\ 
    Q_{3} \big( \theta_{3} ; \boldsymbol{\theta}^{(k)} \big)
    & = \mathbb{E}
    \left[
    \sum_{i=1}^{m} 
    \log f(b_{i})
    \bigg| \ 
    \boldsymbol{y}, \boldsymbol{d}, \boldsymbol{x}, \boldsymbol{z}, \boldsymbol{\theta}^{(k)}
    \right]
\end{align*}

These expectations contain a conditional density of $b_{i}$ given observed covariates which can be expressed as a ratio.
\begin{align}
   f(b_{i} | \boldsymbol{y}_{i}, \boldsymbol{d}_{i}, \boldsymbol{x}_{i}, \boldsymbol{z}_{i})
   =
   \frac{
   f(\boldsymbol{y}_{i} | \boldsymbol{d}_{i}, b_{i}, \boldsymbol{x}_{i}, \boldsymbol{z}_{i})
   f (\boldsymbol{d}_{i} | b_{i}, \boldsymbol{x}_{i}, \boldsymbol{z}_{i})
   f (b_{i})
   }{
   \int 
   f(\boldsymbol{y}_{i} | \boldsymbol{d}_{i}, b_{i}, \boldsymbol{x}_{i}, \boldsymbol{z}_{i})
   f (\boldsymbol{d}_{i} | b_{i}, \boldsymbol{x}_{i}, \boldsymbol{z}_{i})
   f (b_{i})
   d b_{i}
   }.
   \label{Eq:11}
\end{align}
The denominator in \eqref{Eq:11} corresponds to the observed data log-likelihood function, which can be written as:
\begin{align}
    \ell_{obs} (\boldsymbol{\theta}) 
    = \int \sum_{i=1}^{m}
    \log f(\boldsymbol{y}_{i}, \boldsymbol{d}_{i}, b_{i}, \boldsymbol{x}_{i}, \boldsymbol{z}_{i}) d b_{i}.
    \label{Eq:12}
\end{align}
Thus, calculating this integral is necessary for constructing the E-step function \eqref{Eq:10}. However, because the PS matching model has a non-linear form, deriving the exact values of $Q_{l} \big(\boldsymbol{\theta}_{l} ; \boldsymbol{\theta}^{(k)})$ for $l=1,2,3$ is challenging. To address this, we employ approximation methods for each $Q_{l} \big( \boldsymbol{\theta}_{l} ; \boldsymbol{\theta}^{(k)} \big)$ as discussed in Section 3.2.2.

\subsubsection{The Laplace Approximation}


The Laplace approximation is a powerful tool for handling intractable integrals that cannot be computed analytically (Citation?). For instance, in Bayesian statistics, the Laplace approximation is used to approximate the posterior distribution or marginal likelihood function when these are analytically intractable (\citealp{tierney1986accurate}; \citealp{gelman2013bayesian}). This technique is also employed to approximate the integrals involved in the likelihood function for generalized linear mixed-effects models, which may otherwise be intractable \citep{breslow1993approximate}.

As briefly discussed in Section 3.2.1, a significant challenge when using the classical EM algorithm is the inability to obtain closed-form expressions for the integral in the E-step function \eqref{Eq:10}, such as those in \eqref{Eq:11} and \eqref{Eq:12}. To address this challenge, we approximate the $Q$ function using Laplace's method with an extra positive factor. Suppose $g: \mathbb{R}^{p} \to \mathbb{R}$ is a smooth scalar function of $b_{i}$ with a unique minimum at $\tilde{b}_{i}$. If there is an additional factor $h(\cdot)$, which is a smooth and positively valued function, we can employ an alternative Laplace's approximation:
\begin{align}
    \int h(b_{i}) \exp \{ - N g(b_{i}) \} d b_{i}
    \approx
    \left( \frac{2 \pi}{N} \right)^{p/2}
    \frac{
    h(\tilde{b}_{i})
    e^{-N g(\tilde{b}_{i})
    }
    }{
    | H(g) (\tilde{b}_{i}) |^{1/2}
    }
    \label{Eq:13}
\end{align}
where $H(g) (\tilde{b}_{i})$ is a Hessian matrix of $g$ evaluated at $\tilde{b}_{i}$ \citep{butler2007saddlepoint}. Considering that the conditional density \eqref{Eq:11}, choices of $h$ and $g$ for the Laplace's approximation (\ref{Eq:13}) would be of the form:
\begin{align*}
    h(b_{i})
    & = s(b_{i})
    f(\boldsymbol{d}_{i} | b_{i}, \boldsymbol{x}_{i}, \boldsymbol{z}_{i}) \\
    g(b_{i})
    & = -\frac{1}{n} \big\{ 
    \log m(b_{i})
    + \log f(\boldsymbol{y}_{i} | \boldsymbol{d}_{i}, b_{i}, \boldsymbol{x}_{i}, \boldsymbol{z}_{i})
    + \log f(b_{i})
    \big\}
\end{align*}
where $s(\cdot)$ and $m(\cdot)$ are smooth and positive functions. In this study, we can fully obtain the updates for all parameters (M-step) described in Section 3.2.3 by performing following approximations: \textit{Case 1}, \textit{Case 2}, and \textit{Case 3}.

\hfill

\noindent \textit{Case 1.} If we choose $s(b_{i}) = 1$ and $m(b_{i}) = 1$, we will get
\begin{align}
    \int f (\boldsymbol{y}_{i} | \boldsymbol{d}_{i}, b_{i}, \boldsymbol{x}_{i}, \boldsymbol{z}_{i})
    f(\boldsymbol{d}_{i} | b_{i}, \boldsymbol{x}_{i}, \boldsymbol{z}_{i})
    f(b_{i})
    d b_{i}
    \approx
    \sqrt{\frac{2\pi}{N} }
    \frac{
    h (\tilde{b}_{i}) e^{-N g(\tilde{b}_{i})}
    }{
    | H (g) (\tilde{b}_{i})|^{1/2}
    }
    \label{Eq:14}
\end{align}
where the unique minimum $\tilde{b}_{i}$ and the Hessian matrix $H(g)$ are
\begin{align*}
    \tilde{b}_{i}
    & = \left\{ 
    \frac{1}{\sigma^{2}} (\boldsymbol{1} + \omega \boldsymbol{d}_{i})^{\top} (\boldsymbol{1} + \omega \boldsymbol{d}_{i})
    + \frac{1}{\sigma_{b}^{2}}
    \right\}^{-1}
    \left\{ \frac{1}{\sigma^{2}}
    (\boldsymbol{y}_{i} - \tilde{\boldsymbol{x}}_{i} \boldsymbol{\beta})^{\top} (\boldsymbol{1} + \omega \boldsymbol{d}_{i})
    \right\}, \\
    H(g) 
    & = \frac{1}{N} \left\{ \frac{1}{\sigma^{2}} (\boldsymbol{1} + \omega \boldsymbol{d}_{i})^{\top} (\boldsymbol{1} + \omega \boldsymbol{d}_{i}) + \frac{1}{\sigma_{b}^{2}} \right\}.
\end{align*}
The first case calculates the $O(n^{-1})$ approximation of the incomplete-data likelihood \eqref{Eq:12} or the denominator part within the integral as in \eqref{Eq:11}.

\hfill

\noindent \textit{Case 2.} If we choose $s(b_{i}) = 1$ and $m(b_{i}) = e^{t b_{i}}$ for $-c < t < c \ (c > 0)$, we will get an order of $O(n^{-1})$ approximation of the integral,
\begin{align}
    \int e^{t b_{i}} f (\boldsymbol{y}_{i} | \boldsymbol{d}_{i}, b_{i}, \boldsymbol{x}_{i}, \boldsymbol{z}_{i})
    f(\boldsymbol{d}_{i} | b_{i}, \boldsymbol{x}_{i}, \boldsymbol{z}_{i})
    f(b_{i})
    d b_{i}
    & \approx
    \sqrt{\frac{2\pi}{N} }
    \frac{
    h \big( \tilde{b}_{i} (t) \big) e^{-N g(\tilde{b}_{i} (t))}
    }{
    | H (g) (\tilde{b}_{i} (t))|^{1/2}
    }.
    \label{Eq:15}
\end{align}
In this case, the Hessian matrix $H(g)$ is the same with \textit{Case 1}, and the unique minimum is
\begin{align*}
    \tilde{b}_{i} (t)
    = \left\{ 
    \frac{1}{\sigma^{2}} (\boldsymbol{1} + \omega \boldsymbol{d}_{i})^{\top} (\boldsymbol{1} + \omega\boldsymbol{d}_{i})
    +\frac{1}{\sigma_{b}^{2}}
    \right\}^{-1}
    \left\{ 
    t + \frac{1}{\sigma^{2}}
    (\boldsymbol{y}_{i} - \tilde{\boldsymbol{x}}_{i} \boldsymbol{\beta})^{\top} (\boldsymbol{1} + \omega\boldsymbol{d}_{i})
    \right\}.
\end{align*}
The second case accounts for the numerator part of the posterior moment generating function (mgf). That is, the ratio of \textit{Case 1} to \textit{Case 2} will address the approximation of the posterior mgf,
\begin{align}
    \mathbb{E}
    \big[ e^{t b_{i}} | \boldsymbol{y}_{i}, \boldsymbol{d}_{i}, \boldsymbol{x}_{i}, \boldsymbol{z}_{i} \big]
    = \frac{
    \int e^{t b_{i}} f (\boldsymbol{y}_{i} | \boldsymbol{d}_{i}, b_{i}, \boldsymbol{x}_{i}, \boldsymbol{z}_{i})
    f(\boldsymbol{d}_{i} | b_{i}, \boldsymbol{x}_{i}, \boldsymbol{z}_{i})
    f(b_{i})
    d b_{i}
    }{
    \int f (\boldsymbol{y}_{i} | \boldsymbol{d}_{i}, b_{i}, \boldsymbol{x}_{i}, \boldsymbol{z}_{i})
    f(\boldsymbol{d}_{i} | b_{i}, \boldsymbol{x}_{i}, \boldsymbol{z}_{i})
    f(b_{i})
    d b_{i}
    }
    \approx
    \frac{\textit{Case 2}}{\textit{Case 1}}.
    \label{Eq:16}
\end{align}
This approximation of the ratio is justified by \cite{doi:10.1080/01621459.1986.10478240}, which demonstrated that applying the same Laplace approximation to both the numerator and the denominator introduces similar errors, consistent with the typical behavior of Laplace's method. As a result, the error in \eqref{Eq:16} is of order $O(n^{-2})$. Instead of directly approximating the posterior mean $\mathbb{E} \big[ b_{i} | \boldsymbol{y}_{i}, \boldsymbol{d}_{i}, \boldsymbol{x}_{i}, \boldsymbol{z}_{i} \big]$ and posterior variance $Var \big( b_{i} | \boldsymbol{y}_{i}, \boldsymbol{d}_{i}, \boldsymbol{x}_{i}, \boldsymbol{z}_{i} \big)$, we compute these quantities using the approximation in \eqref{Eq:16} as follows.
\begin{align*}
   \mathbb{E} 
   \big( b_{i} | \boldsymbol{y}_{i}, \boldsymbol{d}_{i}, \boldsymbol{x}_{i}, \boldsymbol{z}_{i} \big)
   & = \frac{\partial}{\partial t}
   \log \mathbb{E} 
   \big( e^{t b_{i}} | \boldsymbol{y}_{i}, \boldsymbol{d}_{i}, \boldsymbol{x}_{i}, \boldsymbol{z}_{i} \big)
   \Big|_{t=0} \ \ , \\
   Var 
   \big( b_{i} | \boldsymbol{y}_{i}, \boldsymbol{d}_{i}, \boldsymbol{x}_{i}, \boldsymbol{z}_{i} \big)
   & = \frac{\partial^{2}}{\partial t^{2}}
   \log \mathbb{E} 
   \big( e^{t b_{i}} | \boldsymbol{y}_{i}, \boldsymbol{d}_{i}, \boldsymbol{x}_{i}, \boldsymbol{z}_{i} \big)
   \Big|_{t=0} \ \ .
\end{align*}

This approach is necessary because we cannot set $s(b_{i}) = b_{i}$, given that $h$ in (\ref{Eq:13}) must be positive. This technique was suggested by \cite{azevedo1994laplace} and verified by \cite{tierney1989fully}.


\hfill

\noindent \textit{Case 3.} If we choose $s(b_{i}) =  \log \big\{ 1 + \exp (\boldsymbol{\eta}^{T} \boldsymbol{x}_{ij}^{*} + \xi b_{i}) \big\}$ and $m(b_{i}) = 1$, we obtain an $O(n^{-1})$ approximation of the following integral:
\begin{align*}
   \int 
   \log \big\{ 1 + \exp (\boldsymbol{\eta}^{T} \boldsymbol{x}_{ij}^{*} + \xi b_{i}) \big\}
   f (\boldsymbol{y}_{i} | \boldsymbol{d}_{i}, b_{i}, \boldsymbol{x}_{i}, \boldsymbol{z}_{i})
   f(\boldsymbol{d}_{i} | b_{i}, \boldsymbol{x}_{i}, \boldsymbol{z}_{i})
   f(b_{i})
   d b_{i}
\end{align*}
In \textit{Case 3}, the value of $\tilde{b}_{i}$ and $H(g)$ are the same as those in \textit{Case 1}. The approximated value of the below quantity included in $Q_{2}$
\begin{align*}
    \int \log
    \big\{ 1 + \exp (\boldsymbol{\eta}^{T} \boldsymbol{x}_{ij}^{*} + \xi b_{i}) \big\}
    f(b_{i} | \boldsymbol{y}_{i}, \boldsymbol{d}_{i}, \boldsymbol{x}_{i}, \boldsymbol{z}_{i}) d b_{i}
\end{align*} 
can be obtained by dividing \textit{Case 3} by \textit{Case 1}. These three cases provide approximations for the intractable integral in the $Q$ function \eqref{Eq:10}, and throughout this study, $Q^{*}$ will denote the approximated value of the $Q$ function.





\subsubsection{M-step} 

In the M-step, we aim to find updates of each parameter estimates that maximize (\ref{Eq:9}). For example, maximizing $Q^{*}$ with respect to $\boldsymbol{\beta}$ is equivalent to finding $\boldsymbol{\beta}$ that maximize $Q_{1}^{*}$. Hence, the solution of the equation \eqref{Eq:17} will be the update from $\boldsymbol{\beta}^{(k)}$.
\begin{align}
    \frac{\partial}{\partial \boldsymbol{\beta}}
    Q_{1}^{*} \big( \boldsymbol{\theta}_{1}; \boldsymbol{\theta}^{(k)} \big)
    = 0
    \label{Eq:17}
\end{align}
Denote $\boldsymbol{\beta}^{(k+1)}$ as the update from the current estimate, $\boldsymbol{\beta}^{(k)}$. By solving (\ref{Eq:17}), $\boldsymbol{\beta}^{(k+1)}$ is calculated as
\begin{align}
    \boldsymbol{\beta}^{(k+1)}
    = \left( 
    \sum_{i=1}^{m} \tilde{\boldsymbol{X}}_{i}^{\top} \tilde{\boldsymbol{X}}_{i}
    \right)^{-1}
    \sum_{i=1}^{m} \tilde{\boldsymbol{X}}_{i}^{\top}
    \left\{ 
    \boldsymbol{Y}_{i} - \big( \boldsymbol{1} + \omega^{(k)} \boldsymbol{D}_{i} \big)
    \mu_{i} \big( \boldsymbol{\theta}^{(k)} \big)
    \right\}
    \label{Eq:18}
\end{align}
where $\mu_{i} \big( \boldsymbol{\theta}^{(k)} \big)$ is an approximated value of posterior mean of $b_{i}$, $\mathbb{E} \big( b_{i} | \boldsymbol{y}_{i}, \boldsymbol{d}_{i}, \boldsymbol{x}_{i}, \boldsymbol{z}_{i}, \boldsymbol{\theta}^{(k)} \big)$. Similarly, we can derive the updates for $(\omega, \sigma^{2})$ from the stage 1 and $\sigma_{b}^{2}$ from the stage 3 model. Those are as follows.


\begin{align}
    \omega^{(k+1)}
    & = \left\{ 
    \sum_{i=1}^{m}
    \boldsymbol{1}^{\top} \boldsymbol{D}_{i} \delta_{i} (\boldsymbol{\theta}^{(k)})
    \right\}^{-1}
    \left\{
    \sum_{i=1}^{m}
    \big( \boldsymbol{Y}_{i} - \tilde{\boldsymbol{X}}_{i} \boldsymbol{\beta} \big)^{\top} \boldsymbol{D}_{i}
    \mu_{i} (\boldsymbol{\theta}^{(k)})
    \right\}
    - 1 ,
    \label{Eq:19}
    \\
    \sigma^{2, (k+1)} 
    & = \frac{1}{N}
    \sum_{i=1}^{m} 
    \big( \boldsymbol{y}_{i} - \tilde{\boldsymbol{x}}_{i} \boldsymbol{\beta} \big)^{\top}
    \big( \boldsymbol{y}_{i} - \tilde{\boldsymbol{x}}_{i} \boldsymbol{\beta} \big)
    - \frac{2}{N} \sum_{i=1}^{m} 
    \big( \boldsymbol{y}_{i} - \tilde{\boldsymbol{x}}_{i} \boldsymbol{\beta} \big)^{\top}
    \big( \boldsymbol{1} + \omega \boldsymbol{d}_{i} \big) 
    \mu_{i} (\boldsymbol{\theta}^{(k)})
    \label{Eq:20}
    \\
    & \ \ \ + \frac{1}{N}
    \sum_{i=1}^{m} 
    \big( \boldsymbol{1} + \omega \boldsymbol{d}_{i} \big)^{\top} \big( \boldsymbol{1} + \omega \boldsymbol{d}_{i} \big)
    \delta_{i} (\boldsymbol{\theta}^{(k)}), \nonumber \\
    \sigma_{b}^{2, (k+1)}
    & = \frac{1}{m}
    \sum_{i=1}^{m} 
    \delta_{i} ( \boldsymbol{\theta}^{(k)} )
    \label{Eq:21}.
\end{align}
Here, $\boldsymbol{1}$ is a $n_{i} \times 1$ vector consisting of 1, $N$ is the total number of records ($N = \sum_{i=1}^{m} n_{i}$), and $\delta_{i} \big( \boldsymbol{\theta}^{(k)} \big)$ represents an approximated value of $\mathbb{E} \big( b_{i}^{2} | \boldsymbol{y}_{i}, \boldsymbol{d}_{i}, \boldsymbol{x}_{i}, \boldsymbol{z}_{i}, \boldsymbol{\theta}^{(k)} \big)$. 

While the updates of $(\boldsymbol{\beta}, \omega, \sigma^{2}, \sigma_{b}^{2})$ have closed form as in (\ref{Eq:18}) - (\ref{Eq:21}), there is no explicit solution for the updates of $\boldsymbol{\theta}_{2}$ because these are coefficients in the logistic regression. In this case, a numerical optimization method should be used, and thus we employed Newton Raphson's method. To perform the Newton's method of maximizing $Q_{2}^{*}$ with respect to $\boldsymbol{\theta}_{2}$, deriving the first and the second derivatives of $Q_{2}^{*}$ is required. Let $g(\boldsymbol{\eta}, \xi)$ be the first derivative function of $Q_{2}^{*}$ with respect to $\boldsymbol{\eta}$ and $\xi$. That is, $g(\boldsymbol{\eta}, \xi)$ has a following form 
\begin{align*}
    g (\boldsymbol{\eta}, \xi) 
    = \begin{bmatrix} 
    g_{1} ( \boldsymbol{\eta}, \xi ) \\ g_{2} ( \boldsymbol{\eta}, \xi )
    \end{bmatrix}
\end{align*}
where
\begin{align*}
    g_{1} = g_{1} (\boldsymbol{\eta}, \xi)
    & : = \frac{\partial}{\partial \boldsymbol{\eta}} Q_{2}^{*} (\boldsymbol{\theta}_{2} ; \boldsymbol{\theta}^{(k)}), \\
    g_{2} = g_{2} (\boldsymbol{\eta}, \xi)
    & : = \frac{\partial}{\partial \xi} Q_{2}^{*} (\boldsymbol{\theta}_{2} ; \boldsymbol{\theta}^{(k)}).
\end{align*} 
Next, let $\boldsymbol{J}$ be a Jacobian matrix constituted of the second derivatives,
\begin{align*}
    \boldsymbol{J}
    = \begin{bmatrix}
    \frac{\partial g_{1}}{\partial \boldsymbol{\eta}} & \frac{\partial g_{1}}{\partial \xi} \\
    \frac{\partial g_{2}}{\partial \boldsymbol{\eta}} & \frac{\partial g_{2}}{\partial \xi}
    \end{bmatrix}.
\end{align*}
Then, the Newton's method finding maximizers of $Q_{2}^{*}$ with respect to $\boldsymbol{\theta}_{2}$ would be 
\begin{align}
    \begin{bmatrix}
    \boldsymbol{\eta}_{t+1} \\ \xi_{t+1}
    \end{bmatrix}
    & = \begin{bmatrix}
    \boldsymbol{\eta}_{t} \\ \xi_{t}
    \end{bmatrix}
    - \boldsymbol{J}_{(\boldsymbol{\eta}_{t}, \xi_{t})}^{-1}
    \ g(\boldsymbol{\eta}_{t}, \xi_{t})
    \label{Eq:22}
\end{align}
where $g(\boldsymbol{\eta}_{t}, \xi_{t})$ and $\boldsymbol{J}_{(\boldsymbol{\eta}_{t}, \xi_{t})}$ refer to the previously defined functions evaluated at $\boldsymbol{\eta} = \boldsymbol{\eta}_{t}$ and $\xi=\xi_{t}$. The newton's method (\ref{Eq:22}) will iterate until the change
\begin{align*}
    \left\|
    \begin{bmatrix}
    \boldsymbol{\eta}_{t+1} \\ \xi_{t+1}
    \end{bmatrix}
    - \begin{bmatrix}
    \boldsymbol{\eta}_{t} \\ \xi_{t}
    \end{bmatrix}
    \right\|^{2}
\end{align*}
is less than a sufficiently small value or until the number of iterations approaches a pre-specified value. Once the newton's method stops, the final values will be the updates for $\boldsymbol{\eta}$ and $\xi$. Through the above M-step (\ref{Eq:18}) - (\ref{Eq:22}), we can update the parameter estimates. These EM iterations are repeated until the algorithm converges.

\subsection{Discussion on Convergence}

In general, the EM-type algorithms guarantee a non-decreasing sequence of the observed log-likelihood function with each iteration \citep{dempster1977maximum}. Under regularity conditions, this property ensures that the EM algorithm converges to a stationary point of the likelihood function, which could be a local maximum \citep{wu1983convergence}. Although the Laplacian-Variant EM algorithm might lose this monotonicity due to the combination with Laplace approximation, the convergence of the proposed algorithm remains reliable when the approximations are precise and the initial values are sufficiently close to the global maximum. This reliability stems from the fact that the M-step updates for $(\boldsymbol{\beta}, \omega, \sigma_{b}, \sigma)$, as given by \eqref{Eq:18} - \eqref{Eq:21}, correspond to those in the conventional EM algorithm. Additionally, the M-step updates for $(\boldsymbol{\eta}, \xi)$ are similar to the variant of the EM algorithm proposed by \cite{lange1995gradient} and \cite{steele1996modified}. 


\subsubsection{Initial Value Selection}

Although the EM algorithm is a powerful tool for addressing incomplete data problems, it is highly sensitive to starting points. Depending on the selection of initial values, the algorithm can converge very slowly or potentially converge to local maxima if the likelihood equation has multiple roots \citep{mclachlan2004algorithm}. Therefore, selecting appropriate initial values is crucial for the effective implementation of the EM algorithm. We propose a strategic approach to determine suitable starting points for the algorithm.

Considering the OR model \eqref{Eq:3} and the PS matching model \eqref{Eq:4}, the general strategy involves fitting a linear mixed-effects model and a generalized linear mixed-effects model. To obtain reasonable initial values for $\boldsymbol{\beta}, \omega$, and $\sigma^{2}$, we first fit a linear mixed effects model on the outcome $\boldsymbol{y}_{i}$ using $\tilde{\boldsymbol{x}}_{i}$ and the subject level random effect. The estimated fixed effects and error variances from the fitted linear mixed effects model can serve as initial values for initial values for $\boldsymbol{\beta}$ and $\sigma^{2}$, respectively.

Next, we fit two mixed effects models after separating the data into treated ($D_{ij}=1$) and untreated $(D_{ij}=0$) groups to obtain a good initialization for $\omega$ and $\sigma_{b}^{2}$. Based on the model formulation in \eqref{Eq:2}, the fitted models for the treated and untreated groups are represented by \eqref{Eq:23}) and \eqref{Eq:24}, respectively:
\begin{align}
    (D_{ij} = 1) \ \ \ 
    \boldsymbol{Y}_{i}
    & = ( \boldsymbol{\beta}_{1}^{\top} + \boldsymbol{\beta}_{2}^{\top} ) \boldsymbol{X}_{i}^{*}
    + (1 + \omega) b_{i} + \boldsymbol{\epsilon}_{i}
    \label{Eq:23} 
    \\
    (D_{ij} = 0) \ \ \ 
    \boldsymbol{Y}_{i}
    & = \boldsymbol{\beta}_{1}^{\top} \boldsymbol{X}_{i}^{*} + b_{i} + \boldsymbol{\epsilon}_{i}
    \label{Eq:24}
\end{align}

The estimated variance of the random effects from \eqref{Eq:24} can be used as the initial value of $\sigma_{b}^{2}$. For model \eqref{Eq:23}, the estimated variance of random effects would be close to $(1+\omega)^{2} \sigma_{b}^{2}$. Hence, the initial value for $\omega$ can be obtained by taking the ratio of the estimates from \eqref{Eq:23}) and \eqref{Eq:24}. Finally, we fit the generalized linear mixed effects model (\eqref{Eq:25}) to determine the initial values of $\boldsymbol{\eta}$ and $\xi$:
\begin{align}
    logit P(D_{ij} = 1 | \boldsymbol{x}_{ij}^{*})
    = \boldsymbol{\eta}^{\top} \boldsymbol{x}_{ij}^{*} + b_{i}^{*}
    \label{Eq:25}
\end{align}
Here, $b_{i}^{*}$ represents the subject-level random effect. Similar to the choice of initial values for $\boldsymbol{\beta}$, the estimated fixed effect from (\ref{Eq:25}) can be used as the starting value for $\boldsymbol{\eta}$. Considering the model (\ref{Eq:4}), we can anticipate that the estimate of $Var(b_{i}^{*})$ would be close to $\xi^{2} \sigma_{b}^{2}$. Therefore, we can select $\pm \sqrt{Var(b_{i}^{*})/\sigma_{b}^{2}}$ as a possible initial values for $\xi$. Through simulations, we have confirmed that our strategy yields estimation results and convergence speeds comparable to those achieved when the proposed Laplacian-Variant EM algorithm is initialized from the true values.

\section{Numerical Studies}\label{numerical}

\subsection{Simulation Studies}

In this section, we performed simulation studies to assess the accuracy of estimating model parameters using the proposed Laplacian-Variant EM algorithm. For this purpose, we considered the following OR model:
\begin{align*}
    Y_{ij} 
    = & \beta_{0} +  \beta_{1} Z_{1,i} + \beta_{2} Z_{2,i} + \beta_{3} X_{1, ij} + \beta_{4} X_{2, ij} + \beta_{5} X_{3, ij} \\
    & + D_{ij} \big( 
    \beta_{6} + \beta_{7} Z_{1, i} + \beta_{8} Z_{i, 2}
    + \beta_{9} X_{1, ij} + \beta_{10} X_{2, ij} + \beta_{11} X_{3, ij} \big)
    + (1 + \omega D_{ij}) b_{i} + \epsilon_{ij}
\end{align*}
for $i=1, \ldots, m$ and $j=1, \ldots, n_{i}$ where $\boldsymbol{\beta}=(\beta_{0}, \beta_{1}, \beta_{2}, \beta_{3}, \beta_{4}, \beta_{5}, \beta_{6}, \beta_{7}, \beta_{8}, \beta_{9}, \beta_{10}, \beta_{11})^{\top}$ is set to be $ (-3, 1, 3, -1, -3, 2, 5, 2, 2, -2, -3, 3)^{\top}$. The integrated effect of unmeasured confoundings, denoted as $b_{i}$, is generated from a Normal distribution with a mean 0 and $\sigma_{b} = 1$. The two baseline covariates $Z_{1,i}$ and $Z_{2,i}$ are slightly correlated binary variables with correlation of 0.25, and their marginal probabilities are 0.5 and 0.3, respectively. The three time-variant confounders $X_{1, ij}, X_{2, ij}$, and $X_{3,ij}$ are generated from a Normal distribution with a mean $\boldsymbol{0}$ and an AR(1) covariance structure with a correlation 0.3.  Additionally, in order to address the interaction between unmeasured effect $(b_{i})$ and treatment assignments, the weight $\omega$ is set to be 0.5. The binary treatment variable $D_{ij}$ is generated from the Bernoulli distribution with a probability of the $i$-th unit at time $j$ being treated as follows:
\begin{align*}
    P (D_{ij} = 1 | b_{i}, \boldsymbol{x}_{ij}, \boldsymbol{z}_{i})
    & = \frac{
    \exp (\eta_{0} + \eta_{1} z_{1,i } + \eta_{2} z_{2, i} + \eta_{3} x_{1, ij} + \eta_{4}  x_{2, ij} +\eta_{5}  x_{3, ij} + \xi b_{i})
    }{
    1 + \exp (\eta_{0} + \eta_{1} z_{1,i } + \eta_{2} z_{2, i} + \eta_{3} x_{1, ij} + \eta_{4}  x_{2, ij}+ \eta_{5}  x_{3, ij} + \xi b_{i})
    }
\end{align*}
where the true $\boldsymbol{\eta} = (\eta_{0}, \eta_{1}, \eta_{2}, \eta_{3}, \eta_{4}, \eta_{5})^{\top}$ is set to be $(0.3, -0.3, 0.2, -0.2, 0.2, -0.3)^{\top}$ and true $\xi$ is set to be 0.5 to account for the varying impact of $b_{i}$ between the outcome and treatment assignment.The error term $\epsilon_{ij}$ is generated from a Normal distribution with a mean 0 and $\sigma=0.5$. We employed simulation studies with $m=100, 200$ to compare the estimation accuracy. The number of repetitions was randomly sampled from the set $\{2, 3, \ldots, 10\}$.

Table \ref{Table:1} presents the simulation results based on 200 replications. The estimation results are reported in terms of mean, standard error (S.E.), bias, and root mean squared error (RMSE). From Table \ref{Table:1}, it is evident that the model parameters are estimated accurately. 

\begin{table}[h]
\centering
\caption{Estimation results based on 200 replications}
\vspace{.1in}
\begin{tabular}{cc|lll|lll}
\hline
\multirow{3}{*}{\textbf{Param}} &
  \multirow{3}{*}{\textbf{True}} &
  \multicolumn{3}{c|}{\textbf{nobs=100}} &
  \multicolumn{3}{c}{\textbf{nobs=200}} \\ \cline{3-8} 
             &      & \multicolumn{3}{c|}{\textbf{Estimate}} & \multicolumn{3}{c}{\textbf{Estimate}} \\ \cline{3-8} 
 &
   &
  \textbf{Mean (S.E.)} &
  \textbf{Bias} &
  \textbf{RMSE} &
  \textbf{Mean (S.E.)} &
  \textbf{Bias} &
  \textbf{RMSE} \\ \hline
$\beta_{0}$  & -3   & -2.987 (0.158) & 0.013  & 0.160 & -2.966 (0.108) & 0.034  & 0.113  \\
$\beta_{1}$  & 1    & 1.027 (0.215)  & 0.027  & 0.214 & 1.003 (0.166)  & 0.003  & 0.166  \\
$\beta_{2}$  & 3    & 3.022 (0.254)  & 0.022  & 0.256 & 2.992 (0.159)  & -0.008 & 0.159  \\
$\beta_{3}$  & -1   & -1.001 (0.034) & -0.001 & 0.035 & -0.998 (0.024) & 0.002  & 0.024  \\
$\beta_{4}$  & -3   & -3.001 (0.037) & -0.001 & 0.038 & -3.001 (0.024) & -0.001 & 0.024  \\
$\beta_{5}$  & 2    & 2.001 (0.036)  & 0.001  & 0.037 & 2.000 (0.022)  & 0.000  & 0.022  \\
$\beta_{d}$  & 5    & 5.016 (0.106)  & 0.016  & 0.112 & 5.024 (0.074)  & 0.024  & 0.077  \\
$\beta_{7}$  & 2    & 2.012 (0.140)  & 0.012  & 0.147 & 2.000 (0.104)  & 0.000  & 0.104  \\
$\beta_{8}$  & 2    & 2.007 (0.162)  & 0.007  & 0.165 & 1.988 (0.118)  & -0.012 & 0.118  \\
$\beta_{9}$  & -2   & -1.995 (0.044) & 0.005  & 0.049 & -2.002 (0.034) & -0.002 & 0.034  \\
$\beta_{10}$ & -3   & -3.004 (0.044) & -0.004 & 0.046 & -2.998 (0.032) & 0.002  & 0.032  \\
$\beta_{11}$ & 3    & 3.002 (0.047)  & 0.002  & 0.049 & 3.000 (0.033)  & 0.000  & 0.033  \\
$\omega$     & 0.5  & 0.51  (0.063)  & 0.01   & 0.089 & 0.508 (0.049)  & 0.008  & 0.050  \\
$\sigma$     & 0.5  & 0.495 (0.015)  & -0.005 & 0.016 & 0.497 (0.012)  & -0.003 & 0.012  \\
$\eta_{0}$   & 0.3  & 0.328 (0.139)  & 0.028  & 0.203 & 0.327 (0.106)  & 0.027  & 0.109  \\
$\eta_{1}$   & -0.3 & -0.307 (0.207) & -0.007 & 0.282 & -0.305 (0.153) & -0.005 & 0.152  \\
$\eta_{2}$   & 0.2  & 0.218 (0.248)  & 0.018  & 0.354 & 0.195 (0.143)  & -0.005 & 0.143  \\
$\eta_{3}$   & -0.2 & -0.210 (0.088) & -0.010 & 0.091 & -0.208 (0.059) & -0.008 & 0.060  \\
$\eta_{4}$   & 0.2  & 0.199 (0.083)  & -0.001 & 0.094 & 0.207 (0.065)  & 0.007  & 0.066  \\
$\eta_{5}$   & -0.3 & -0.316 (0.089) & -0.016 & 0.096 & -0.301 (0.061) & -0.001 & 0.061  \\
$\xi$        & 0.5  & 0.513 (0.109)  & 0.013  & 0.150 & 0.506 (0.073)  & 0.006  & 0.073  \\
$\sigma_{b}$ & 1    & 0.981 (0.082)  & -0.019 & 0.090 & 0.992 (0.061)  & -0.008 & 0.061  \\
ATE          & 6.6  & 6.632 (0.181)  & 0.032  & 0.183 & 6.631 (0.132)  & 0.031  & 0.135  \\ \hline
\end{tabular}
\label{Table:1}      
\end{table}

\subsection{Application to CitieS-Health Barcelona Panel Study}


In this section, we apply the proposed method to air pollution and mental health data from CitieS-Health Barcelona Panel Study, which was carried out by \cite{gignac_florence_2022_6503022}. The dataset comprises 3,333 observations from 286 distinct participants, collected in Barcelona, Spain. It encompasses various environmental variables, including levels of air pollution (specifically PM2.5, black carbon, and nitrogen dioxide) and meteorological variables such as mean temperature, humidity, and precipitation. Additionally, self-reported survey results related to mental health, physical activity, dietary habits, and smoking behavior are recorded. Furthermore, the dataset includes results from the Stroop test, a psychological assessment tool used to gauge cognitive processing speed and selective attention, which serves as an indicator of mental health. During this test, participants are presented with a list of color names written in different colors. The participant's task is to name the ink color of each word as quickly and accurately as possible, regardless of the actual word. Assessing Stroop test performance involves analyzing the participants' response times and accuracy under different conditions, including congruent and incongruent conditions. In the congruent condition, color words appear to the participants in colors that match the actual color, while in the incongruent condition, color words that do not match the actual colors are presented.


\begin{table}[h]
\centering
\caption{Descriptive statistics for the continuous variable.}
\vspace{.1in}
\begin{tabular}{lccl}
\hline
           & Mean (SD)         & Range            & Type        \\ \hline
age        & 38.2  (12.3)      & (18, 76)         & Integer     \\ 
stress     & 4.242 (2.498)     & (0, 10)          & Integer     \\
noise      & 2.124 (3.450)     & (0, 20.005)      & Continuous  \\
mean\_temp & 16.377 (3.191)    & (7.598, 25.956)  & Continuous  \\
humid      & 65.843 (13.653)   & (31.313, 97.688) & Continuous  \\ \hline
\label{Table:2}
\end{tabular}
\end{table}

In this study, we considered the $z$-scores of Stroop test performance as our outcome variable ($\boldsymbol{Y}$). We created a new binary treatment variable ($\boldsymbol{D}$), named \texttt{pm2.5\_good}, which takes the value 1 if the PM2.5 level in Barcelona is less than or equal to 12 \si{\micro\gram}/$m^{3}$, and 0 otherwise. This criterion aligns with the annual average standard for PM2.5 levels. After eliminating highly correlated variables, we selected participants' gender (\texttt{female}: 1 for female, 0 for male), \texttt{age} in years, and \texttt{education} (1 for university-level, 0 for below university level) as baseline covariates ($\boldsymbol{Z}$). Additionally, we considered \texttt{stress} levels (0 to 10 integers), hours of noise over 65 dB in a day (\texttt{noise}), mean temperature ($^{\circ} C$) over 24 hours (\texttt{mean\_temp}), and humidity over 24 hours (\texttt{humid}) as time-varying confounders ($\boldsymbol{X}$). Descriptive statistics for the continuous variables are summarized in Table \ref{Table:2}.  Lastly, in this dataset, the integrated effects of unmeasured confounding variables $(b_{i})$ may include factors such as individual characteristics, proximity to green or blue areas, urban transportation intensity, or the presence of nearby factories.

\begin{table}[h]
\centering
\caption{Estimates of Average Treatment Effect}
\vspace{.05in}
\begin{tabular}{lcccccc}
\hline
Estimate & Mean   & Std. Err & 95\% C.I.        & 90\% C.I.       & P-value  \\ \hline
0.043    & 0.043  & 0.024    & (-0.002, 0.091)  & (0.004, 0.082)  & 0.078    \\ \hline
\label{Table:3} 
\end{tabular} \\ 
\begin{flushleft}
Note) Estimate means point estimates from the Laplacian Variant EM algorithm. Mean, Std. Err, C.I., and P-value represent a bootstrap mean, standard error, confidence interval, and p-value, respectively.
\end{flushleft}
\normalsize
\end{table}

Table \ref{Table:3} presents the estimates of the ATE of good PM2.5 levels on cognitive test performance, as derived from the proposed Laplacian-Variant EM algorithm. Given that we employed clustered or multilevel modeling using panel data, we utilized the cluster bootstrap method as a resampling technique (\citealp{field2007bootstrapping}; \citealp{leeden2008resampling}). The results in Table 3 suggest that better air quality, as indicated by good PM2.5 levels, has a causal effect on modest improvements in cognitive test performance. This effect is statistically significant at the 90\% confidence level.

\begin{figure}[h]
   \centering
   \includegraphics[width=\textwidth]{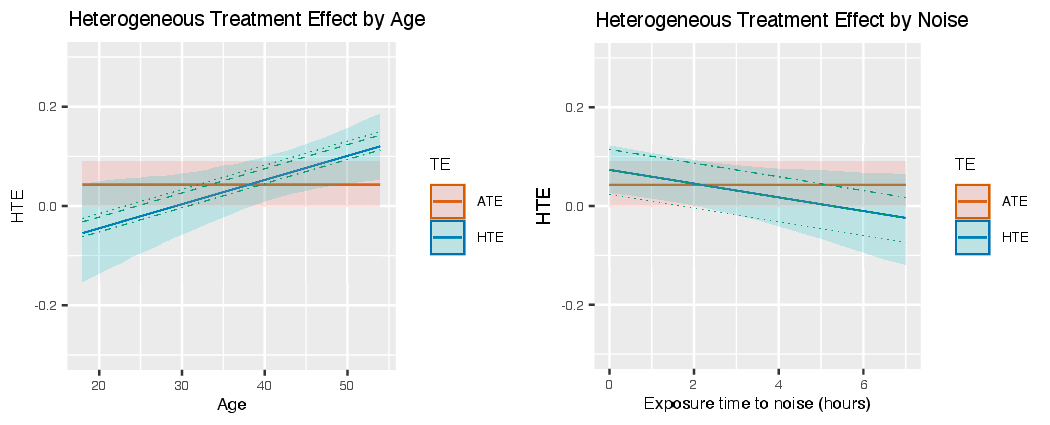}
   \caption{Heterogeneous Treatment Effect by age (Left) and exposure time to noise}
   \label{Fig.1}
\end{figure}



The left plot in Figure \ref{Fig.1} illustrates how the heterogeneous treatment effect (HTE) changes with participants' age, conditioned on the sample mean (solid line), median (dashed line), first quartile (dotted line), and third quartile (dot-dash line) of exposure time to noise. The plot shows an increasing trend with age across all levels of noise exposure. This indicates that as participants get older, the beneficial effect of good PM2.5 levels on cognitive performance tends to increase. The confidence band for the HTE excludes zero for participants older than approximately 40 years. This suggests that, for older participants, the treatment effect of good PM2.5 levels on cognitive performance is statistically significant. For participants older than 50 years, the confidence band for the HTE is beyond the ATE line (red line). This indicates that the treatment effect for these participants is significantly different and higher than the average treatment effect, highlighting a stronger benefit from good PM2.5 levels on cognitive performance for this age group.


The right plot in Figure \ref{Fig.1} examines how the HTE changes with exposure time to noise, given the sample mean (solid line), median (dashed line), first quartile (dotted line), and third quartile (dot-dash line) of participants' age. This plot reveals a decreasing trend with increased exposure time to noise, suggesting that the cognitive benefits of good PM2.5 levels diminish as noise exposure increases. The confidence band for the HTE excludes zero for exposure times less than approximately 4 hours. This suggests that, with shorter noise exposure, the treatment effect of good PM2.5 levels on cognitive performance is statistically significant. The confidence band for the HTE does not exclude the ATE (red line) across the range of exposure times to noise shown. This indicates that the ATE is a reasonable summary of the treatment effect across all noise exposure times, and the variability in the HTE does not significantly deviate from the average treatment effect.

\begin{table}[h]
\centering
\caption{Estimates of $\boldsymbol{\beta}$ and $\omega$ in the OR}
\vspace{.05in}
\begin{tabular}{lcccccc}
\hline
Variable          & Estimate & Mean   & Std. Err & 95\% C.I.        & P-value  \\ \hline
Intercept         & 0.985    & 0.979  & 0.182    & (0.617, 1.337)   & 0.000    \\
age               & -0.032   & -0.032 & 0.003    & (-0.038, -0.027) & 0.000    \\
female            & 0.015    & 0.016  & 0.085    & (-0.157, 0.186)  & 0.844    \\
education         & 0.235    & 0.237  & 0.095    & (0.054, 0.431)   & 0.015    \\
stress            & -0.012   & -0.012 & 0.006    & (-0.023, 0)      & 0.043    \\
noise             & 0.015    & 0.015  & 0.005    & (0.005, 0.026)   & 0.006    \\
mean\_temp        & 0.003    & 0.003  & 0.007    & (-0.01, 0.018)   & 0.595    \\
humid             & -0.001   & -0.001 & 0.001    & (-0.003, 0.001)  & 0.379    \\
pm2.5\_good       & -0.113   & -0.106 & 0.082    & (-0.276, 0.058)  & 0.163    \\
pm2.5\_good*age   & 0.005    & 0.005  & 0.002    & (0.001, 0.009)   & 0.014    \\
pm2.5\_good*noise & -0.014   & -0.014 & 0.007    & (-0.029, 0)      & 0.053    \\ 
$\omega$          & 0.085    & 0.085  & 0.048    & (-0.004, 0.188)  & 0.07    \\ \hline
\label{Table:4} 
\end{tabular} \\ 
\begin{flushleft}
Note) Estimate means point estimates from the Laplacian Variant EM algorithm. Mean, Std. Err, C.I., and P-value represent a bootstrap mean, standard error, confidence interval, and p-value, respectively.
\end{flushleft}
\normalsize
\end{table}

Table \ref{Table:4} shows the estimate of $\boldsymbol{\beta}$ coefficients from the proposed method. Specifically, $\omega$ is estimated at 0.085 (p-value = 0.07), indicating a slight positive interaction between unmeasured effect$(b_{i})$ and the treatment assignment$(d_{ij})$. This implies that the impact of unmeasured confounding variables on Stroop test performance is slightly greater in the group exposed to good PM2.5 level compared to those exposed to worse air pollution conditions. To determine which factors are related to the treatment assignment, we performed a $\chi^{2}$ test using the bootstrap results.

Based on the results in Table \ref{Table:5}, we can conclude that environmental factors—such as exposure time to noise, mean temperature, and humidity over 24 hours—are strongly related to the PM2.5 concentration level. Interestingly, at least one of individual factors such as participants' age, gender, highest education, and stress level are also associated with air quality, although their impact is relatively smaller compared to the environmental factors. Conversely, the integrated effect of unmeasured confounding variables is not associated with the air pollution condition. This suggests that the impact of unmeasured factors on cognitive test scores differs from their impact on exposure to better air quality. 

\begin{table}[h]
\centering
\caption{$\chi^{2}$ Test Results}
\vspace{.05in}
\begin{tabular}{lccc}
\hline
Factor          & Statistic & DF  & P-value \\ \hline
Individual      & 11.933    &  4  & 0.018   \\
Environmental   & 163.460   &  3  & 0.000   \\
Unmeasured      & 0.124     &  1  & 0.725   \\ \hline
\label{Table:5} 
\end{tabular}
\end{table}

Table \ref{Table:6} provides a detailed summary of the estimation results for $\boldsymbol{\eta}$ and $\xi$ in the PS matching model. Additionally, Table \ref{Table:7} shows that the estimate of $\sigma$ and $\sigma_{b}$ are 0.653 and 0.564, respectively. To test for the presence of unmeasured confounding effects $(H_{0}: \sigma_{b} = 0 \ \text{vs.} \ H_{a}: \sigma_{b} > 0)$, we perform a log-likelihood ratio test (LRT), with results summarized in Table \ref{Table:8}. The LRT result shows that $\sigma_{b}$ is significantly greater than zero, indicating that incorporating $b_{i}$ substantially improves model performance.

\begin{table}[h]
\centering
\caption{Estimates of $\boldsymbol{\eta}$ and $\xi$ in the PS modeling}
\vspace{.05in}
\begin{tabular}{lcccccc}
\hline
Variable          & Estimate & Mean   & Std. Err & 95\% C.I.        & P-value  \\ \hline
Intercept         & -0.135   & -0.133 & 0.105    & (-0.344, 0.062)  & 0.200    \\
age               & 0.029    & 0.030  & 0.041    & (-0.049, 0.114)  & 0.461    \\
female            & 0.055    & 0.052  & 0.080    & (-0.109, 0.205)  & 0.487    \\
education         & -0.096   & -0.098 & 0.100    & (-0.298, 0.098)  & 0.325    \\
stress            & -0.101   & -0.102 & 0.039    & (-0.18, -0.026)  & 0.008    \\
noise             & -0.004   & -0.005 & 0.039    & (-0.083, 0.072)  & 0.939    \\
mean\_temp        & 0.006    & 0.004  & 0.035    & (-0.065, 0.073)  & 0.869    \\
humid             & -0.427   & -0.425 & 0.037    & (-0.5, -0.356)   & 0.000    \\
$\xi$             & 0.029    & 0.032  & 0.085    & (-0.128, 0.2)    & 0.738    \\ \hline
\label{Table:6} 
\end{tabular} \\
\begin{flushleft}
Note) Estimate means point estimates from the Laplacian Variant EM algorithm. Mean, Std. Err, C.I., and P-value represent a bootstrap mean, standard error, confidence interval, and p-value, respectively.
\end{flushleft}
\normalsize
\end{table}

\begin{table}[h]
\centering
\caption{Estimate of $\sigma$ and $\sigma_{b}$}
\vspace{.05in}
\begin{tabular}{lcccc}
\hline
              & Estimate &  Mean  & Std.Err & 95\% C.I. \\ \hline
$\sigma$      & 0.653    &  0.652 & 0.015   & (0.621, 0.680) \\
$\sigma_{b}$  & 0.564    &  0.559 & 0.029   & (0.501, 0.616) \\ \hline
\label{Table:7}
\end{tabular} \\
\begin{flushleft}
Note) Estimate means point estimates from the Laplacian Variant EM algorithm. Mean, Std. Err, C.I. represent a bootstrap mean, standard error, and confidence interval, respectively.
\end{flushleft}
\normalsize
\end{table}

\begin{table}[h]
\centering
\caption{Likelihood Ratio Test for $\sigma_{b}$}
\vspace{.05in}
\begin{tabular}{lcccc}
\hline
                         & LogLik     &  DF & $\chi^{2}$ Statistic  & P-value \\ \hline
$H_{0}: \sigma_{b} = 0$  & -4256.787  &  12 &                       &         \\
$H_{a}: \sigma_{b} > 0$  & -3163.900  &  14 & 2185.773              & 0.000   \\ \hline
\label{Table:8}
\end{tabular} \\
\begin{flushleft}
Note) LogLik represents the log-likelihood value of the OR function, while DF denotes the number of parameters under each hypothesis. The $\chi^{2}$ Statistic is calculated as twice the difference between the log-likelihood values under each hypothesis.\end{flushleft}
\normalsize
\end{table}

Next, we perform a sensitivity analysis to justify the presence of the effect of unmeasured confounding variables. In the analysis, we compare the conditional log-likelihood, defined as:

\begin{align*}
   \ell_{cond} ( \boldsymbol{\theta} )
        & = \sum_{i=1}^{m} \sum_{j=1}^{n_{i}}
        \log f (y_{ij}, d_{ij} |b_{i}, \boldsymbol{x}_{ij}, \boldsymbol{z}_{i}, \boldsymbol{\theta} ) \\
        & = \sum_{i=1}^{m} \sum_{j=1}^{n_{i}}
        \left\{ 
        \log f(y_{ij} | d_{ij}, b_{i}, \boldsymbol{x}_{ij}, \boldsymbol{z}_{i}, \boldsymbol{\theta}_{1} )
        + \log f(d_{ij} | b_{i}, \boldsymbol{x}_{ij}, \boldsymbol{z}_{i},\boldsymbol{\theta}_{2})
        \right\}.
\end{align*}

In the full model, $b_{i}$ is incorporated, while in the reduced model, the presence of unmeasured effects is ignored by treating $b_{i}$ as 0. Therefore, the conditional log-likelihood functions for the full and reduced models can be written as:
\begin{align*}
   \ell_{cond}^{F} (\hat{\boldsymbol{\theta}})
   & = \sum_{i=1}^{m} \sum_{j=1}^{n_{i}}
   \left\{ 
   \log f(y_{ij} | d_{ij}, \boldsymbol{x}_{ij}, \boldsymbol{z}_{i}, \hat{b}_{i}, \hat{\boldsymbol{\theta}}_{1} )
   + \log f(d_{ij} | \boldsymbol{x}_{ij}, \boldsymbol{z}_{i}, \hat{b}_{i}, \hat{\boldsymbol{\theta}}_{2})
   \right\}, \\
   \ell_{cond}^{R} (\tilde{\boldsymbol{\theta}})
   & = \sum_{i=1}^{m} \sum_{j=1}^{n_{i}}
   \left\{ 
   \log f(y_{ij} | d_{ij}, \boldsymbol{x}_{ij}, \boldsymbol{z}_{i}, \tilde{\boldsymbol{\theta}}_{1} )
   + \log f(d_{ij} | \boldsymbol{x}_{ij}, \boldsymbol{z}_{i}, \tilde{\boldsymbol{\theta}}_{2})
   \right\}
\end{align*}
where $\hat{\boldsymbol{\theta}}$ indicates the parameter estimates from the proposed Laplacian-Variant EM algorithm and $\tilde{\boldsymbol{\theta}}$ represents the parameter estimates from the reduced model. 

\begin{table}[h]
\centering
\caption{Comparison of the Conditional log-likelihood (CLL)}
\vspace{.05in}
\begin{tabular}{lc}
\hline
Model                 & CLL \\ \hline
Full                  & -5372.546 \\
Full - (age)          & -5729.445 \\
Full - (age \& humid) & -5800.598 \\
Reduced $(b_{i} = 0)$ & -6466.031 \\ \hline
\label{Table:9}
\end{tabular}
\end{table}
Table \ref{Table:9} summarizes the values of the conditional log-likelihood for the full model (Full), models with reduced observed covariates (Full - (age), Full - (age \& humid)), and the model that ignores the effect of unmeasured confounding variables (Reduced). The result indicates that the inclusion of the effect of unmeasured confounding variables $(b_{i})$ significantly improves the model's fit, as evidenced by the much higher CLL value for the full model compared to the reduced model. This finding suggests that the effect of unmeasured confounding variables is stronger than that of the observed covariates, thereby justifying the incorporation of $b_{i}$ in the analysis. This is consistent with the understanding that cognitive performance can be heavily influenced by individual characteristics, such as genetic factors.

\section{Discussion}


In this study, we developed a new mixed-effects joint modeling method to estimate the causal effect of treatment or exposure on outcomes, accounting for both observed covariates and unmeasured confounding variables. 
We proposed a
Laplacian-Variant EM algorithm to estimate the parameters. Our model allows the interaction effects of the treatment and unobserved confounders on the outcome, as well as a varying effect of unmeasured confounders on the outcome and treatment assignment. This provides a robust method for estimating treatment effects in complex data structures, such as those found in environmental health studies.

The results, summarized in Table \ref{Table:3}, indicate that good air quality, defined as PM2.5 levels less than or equal to 12 \si{\micro\gram}/$m^{3}$, has a modest but statistically significant positive effect on cognitive test performance. This effect was particularly evident at the 90\% confidence level, underscoring the relevance of maintaining low PM2.5 levels to support cognitive health. Notably, as shown in Figure \ref{Fig.1}, older participants experience greater cognitive benefits from good PM2.5 levels compared to younger participants, potentially due to increased sensitivity to air quality changes with age. Additionally, longer exposure to noise appears to diminish the cognitive benefits of good PM2.5 levels, likely because noise interferes with cognitive function, mitigating the positive impact of improved air quality.

The estimates of $\omega$ (Table \ref{Table:4}) reveal a slight positive interaction between unmeasured effects and treatment assignment, suggesting that unmeasured factors have a greater impact on cognitive performance under good air quality conditions. Additionally, the inclusion of unmeasured confounding variables in the model is supported by the LRT (Table \ref{Table:8}) and a comparison of conditional log-likelihood values (Table \ref{Table:9}), both indicating that ignoring these unmeasured effects would result in a substantial loss of model fit.

These findings contribute to the growing body of evidence on the impact of air quality on cognitive function. By employing a novel statistical method that integrates both observed and unmeasured confounders, we were able to derive more reliable estimates of the true causal effect of PM2.5 levels on cognitive performance. The slight but statistically significant improvement in cognitive function associated with better air quality emphasizes the importance of environmental policies aimed at reducing air pollution, not only for physical health but also for cognitive well-being. The methodology developed here has broader applications beyond environmental health studies. The Laplacian-Variant EM algorithm can be adapted to other contexts where unmeasured confounding is a concern, providing a valuable tool for causal inference in complex data settings.

However, our use of the annual average standard for PM2.5 exposure (due to a limited number of short-term exposure records exceeding the 24-hour average standard of 35 \si{\micro\gram}/$m^{3}$) may have contributed to the moderate estimated average treatment effect and overlapping confidence intervals for ATE and HTE. This limitation underscores the need for more precise data to better understand the short-term impacts of air quality on cognitive performance. Additionally, the study's findings on the interaction between unmeasured factors and air quality suggest that further research is needed to identify and understand these latent variables. This could involve more detailed data collection or the application of advanced machine learning techniques to uncover hidden patterns in the data.

In conclusion, this study not only demonstrates the cognitive benefits of good air quality but also advances the methodological toolkit for analyzing complex causal relationships in the presence of unmeasured confounding variables. Future studies could build on this work by conducting sensitivity analyses or developing more flexible models for both OR and PS matching.

\newpage

\bibliographystyle{apalike}

\bibliography{references}

\end{document}